\documentclass[aps,prl,longbibliography,superscriptaddress,reprint]{revtex4-2}

\usepackage{graphicx}
\usepackage{amsmath}
\usepackage{multirow} 
\usepackage{physics} 
\usepackage{braket}
\usepackage{times}
\usepackage{gensymb}
\usepackage{hyperref}
\hypersetup{colorlinks=true, citecolor=blue, urlcolor=blue, linkcolor=blue}

\begin{document} 

\title{Multi-parameter quantum metrology with stabilized multi-mode squeezed state}

\author{Yue Li}\email{Equal contribution.}
\affiliation{CAS Key Laboratory of Microscale Magnetic Resonance and School of Physical Sciences, University of Science and Technology of China, Hefei 230026, China}
\affiliation{CAS Center for Excellence in Quantum Information and Quantum Physics, University of Science and Technology of China, Hefei 230026, China}

\author{Xu Cheng}\email{Equal contribution.}
\affiliation{CAS Key Laboratory of Microscale Magnetic Resonance and School of Physical Sciences, University of Science and Technology of China, Hefei 230026, China}
\affiliation{CAS Center for Excellence in Quantum Information and Quantum Physics, University of Science and Technology of China, Hefei 230026, China}
\affiliation{Hefei National Laboratory, University of Science and Technology of China, Hefei 230088, China}

\author{Lingna Wang}\email{Equal contribution.}
\affiliation{Department of Mechanical and Automation Engineering, The Chinese University of Hong Kong, Shatin, Hong Kong}

\author{Xingyu Zhao}
\affiliation{CAS Key Laboratory of Microscale Magnetic Resonance and School of Physical Sciences, University of Science and Technology of China, Hefei 230026, China}
\affiliation{CAS Center for Excellence in Quantum Information and Quantum Physics, University of Science and Technology of China, Hefei 230026, China}
\affiliation{Hefei National Laboratory, University of Science and Technology of China, Hefei 230088, China}

\author{Waner Hou}
\affiliation{CAS Key Laboratory of Microscale Magnetic Resonance and School of Physical Sciences, University of Science and Technology of China, Hefei 230026, China}
\affiliation{CAS Center for Excellence in Quantum Information and Quantum Physics, University of Science and Technology of China, Hefei 230026, China}

\author{Yi Li}
\affiliation{CAS Key Laboratory of Microscale Magnetic Resonance and School of Physical Sciences, University of Science and Technology of China, Hefei 230026, China}
\affiliation{CAS Center for Excellence in Quantum Information and Quantum Physics, University of Science and Technology of China, Hefei 230026, China}

\author{Kamran Rehan}
\affiliation{CAS Key Laboratory of Microscale Magnetic Resonance and School of Physical Sciences, University of Science and Technology of China, Hefei 230026, China}
\affiliation{CAS Center for Excellence in Quantum Information and Quantum Physics, University of Science and Technology of China, Hefei 230026, China}

\author{Mingdong Zhu}
\affiliation{CAS Key Laboratory of Microscale Magnetic Resonance and School of Physical Sciences, University of Science and Technology of China, Hefei 230026, China}
\affiliation{CAS Center for Excellence in Quantum Information and Quantum Physics, University of Science and Technology of China, Hefei 230026, China}

\author{Lin Yan}
\affiliation{CAS Key Laboratory of Microscale Magnetic Resonance and School of Physical Sciences, University of Science and Technology of China, Hefei 230026, China}
\affiliation{CAS Center for Excellence in Quantum Information and Quantum Physics, University of Science and Technology of China, Hefei 230026, China}

\author{Xi Qin}
\affiliation{CAS Key Laboratory of Microscale Magnetic Resonance and School of Physical Sciences, University of Science and Technology of China, Hefei 230026, China}
\affiliation{CAS Center for Excellence in Quantum Information and Quantum Physics, University of Science and Technology of China, Hefei 230026, China}
\affiliation{Hefei National Laboratory, University of Science and Technology of China, Hefei 230088, China}

\author{Xinhua Peng}
\affiliation{CAS Key Laboratory of Microscale Magnetic Resonance and School of Physical Sciences, University of Science and Technology of China, Hefei 230026, China}
\affiliation{CAS Center for Excellence in Quantum Information and Quantum Physics, University of Science and Technology of China, Hefei 230026, China}
\affiliation{Hefei National Laboratory, University of Science and Technology of China, Hefei 230088, China}

\author{Haidong Yuan} \email{hdyuan@mae.cuhk.edu.hk}
\affiliation{Department of Mechanical and Automation Engineering, The Chinese University of Hong Kong, Shatin, Hong Kong}

\author{Yiheng Lin} \email{yiheng@ustc.edu.cn}
\affiliation{CAS Key Laboratory of Microscale Magnetic Resonance and School of Physical Sciences, University of Science and Technology of China, Hefei 230026, China}
\affiliation{CAS Center for Excellence in Quantum Information and Quantum Physics, University of Science and Technology of China, Hefei 230026, China}
\affiliation{Hefei National Laboratory, University of Science and Technology of China, Hefei 230088, China}

\author{Jiangfeng Du} \email{djf@ustc.edu.cn}
\affiliation{CAS Key Laboratory of Microscale Magnetic Resonance and School of Physical Sciences, University of Science and Technology of China, Hefei 230026, China}
\affiliation{CAS Center for Excellence in Quantum Information and Quantum Physics, University of Science and Technology of China, Hefei 230026, China}
\affiliation{Hefei National Laboratory, University of Science and Technology of China, Hefei 230088, China}
\affiliation{Institute of Quantum Sensing and School of Physics, Zhejiang University, Hangzhou 310027, China}

\begin{abstract}
Squeezing a quantum state along a specific direction has long been recognized as a crucial technique for enhancing the precision of quantum metrology by reducing parameter uncertainty. However, practical quantum metrology often involves the simultaneous estimation of multiple parameters, necessitating the use of high-quality squeezed states along multiple orthogonal axes to surpass the standard quantum limit
for all relevant parameters. In addition, a temporally stabilized squeezed state can provide an event-ready probe for parameters, regardless of the initial state, and robust to the timing of the state preparation process once stabilized. In this work, we generate and stabilize a two-mode squeezed state along two secular motional modes in a vibrating trapped ion with reservoir engineering, despite starting from a thermal state of the motion. Leveraging this resource, we demonstrate an estimation of two simultaneous collective displacements along the squeezed axes, achieving improvements surpassing the classical limit by up to 6.9(3) and 7.0(3) decibels (dB), respectively. Our demonstration can be readily scaled to squeezed states with even more modes. The practical implications of our findings span a wide range of applications, including quantum sensing, quantum imaging, and various fields that demand precise measurements of multiple parameters.
\end{abstract}

\maketitle 
By squeezing a quantum state along a specific direction, it is possible to significantly decrease the uncertainty associated with a relevant parameter, leading to a precision beyond the standard quantum limit\cite{Advances_metrology}. This concept has found widespread applications in quantum metrology, enabling improved precision in the measurement of various physical quantities, including displacement, phase, and frequency of a system\cite{Advances_metrology,Quantum_dense,gravitational_wave,Quantum_Enhanced_Measurements,Quantum_amplification,Spin_squeezing,parametric_amplifier,Multiparameter_squeezing,Parity_Detection,LuMetrology,Optomechanical}. The utilization of squeezed states has been particularly notable in the field of gravitational wave detection, exemplified by the Laser Interferometer Gravitational-Wave Observatory (LIGO)\cite{LIGO}, where squeezed states have been utilized to enhance the gravitational wave detection capabilities.

\begin{figure}[!ht]
    \centering
    \includegraphics[width=\columnwidth]{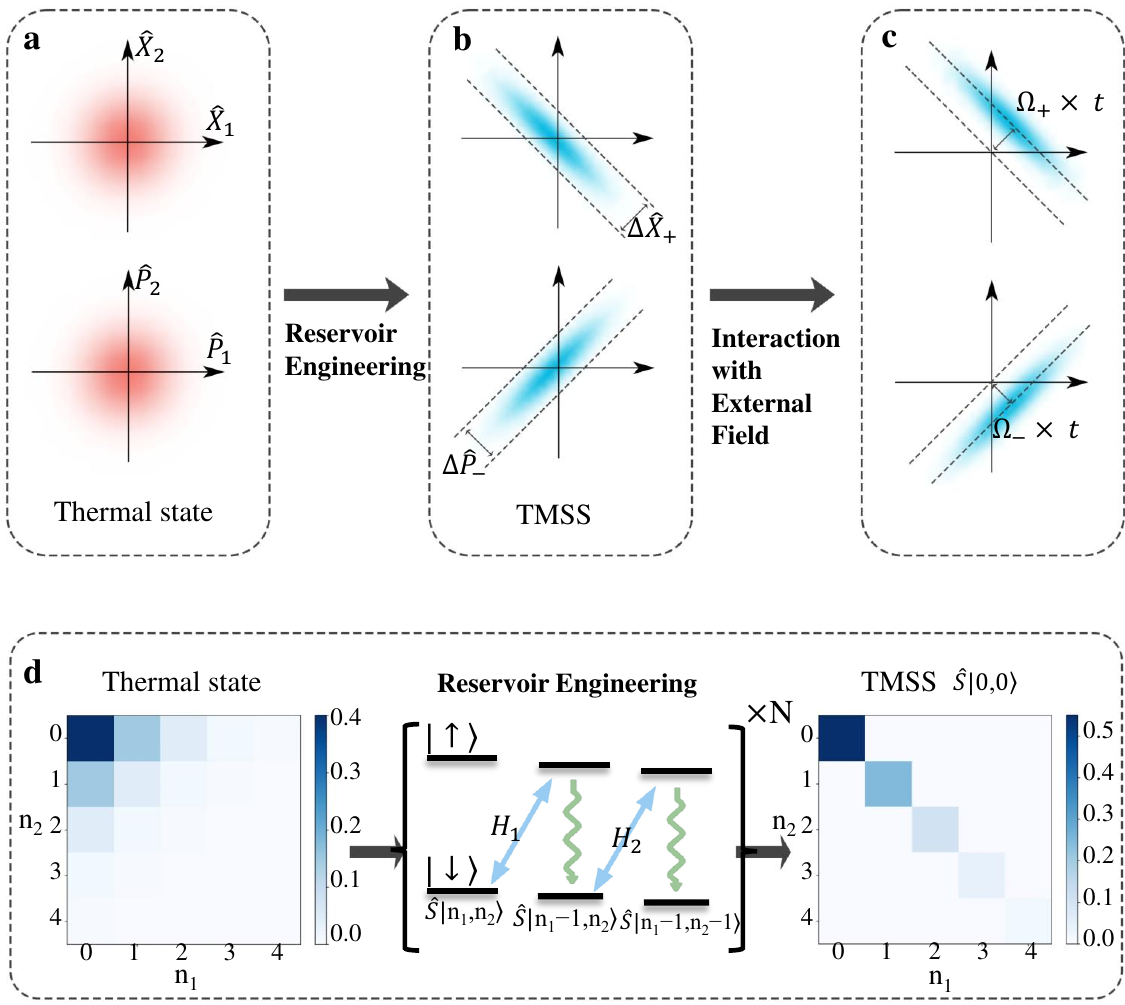}    \caption{\textbf{Conceptual illustration of two-mode squeezed state and simultaneous estimation of two parameters.} Panel \textbf{a},\textbf{b},\textbf{c} display the phase space distribution of the Wigner function. In our experiment, we start with a thermal state after Doppler cooling and EIT cooling (panel \textbf{a}). Subsequently, we apply reservoir engineering, which produces a stable output state regardless of the initial state, generating the desired two-mode squeezed state, denoted as TMSS (panel \textbf{b}). TMSS is squeezed along two different axis $\hat{X}_+$ and $\hat{P}_-$.   Afterward, TMSS can interact with multiple external fields for amplitude ($\Omega_{\pm}$) measurement along the squeezed axes, resulting in enhanced precision(panel \textbf{c}). In panel \textbf{d}, we show the reservoir engineering process converting a thermal state to the desired TMSS state. In Fock basis, we show the population of the input and output state of the process, where the two modes of the desired TMSS output contains superpositions of equal Fock states and is thus entangled.}
\end{figure}

While squeezed states have traditionally been studied and utilized for enhancing the precision of estimating a single parameter, there are many practical scenarios, such as quantum imaging and quantum sensing\cite{PhysRevX.6.031033, albarelli2020perspective},
where multiple parameters come into play. In these situations, conventional methods that use states squeezed in a single direction may fail to provide precision improvement for the estimation of all parameters simultaneously.

One promising approach for exploring the potential of squeezed states in multiple parameter estimation is the use of multi-mode squeezed states\cite{CV, Squeezed_Light}. These states, which exhibit entanglement across various modes, can be squeezed along multiple axes simultaneously. While two-mode squeezed state (TMSS) has been generated previously with light and mechanical
oscillators\cite{ Parity_Detection, LuMetrology, multiphotonMetrology,laser_interferometers,metrology_TMSS,massive,micromechanical,quantum_memory}, 
the simultaneous estimation of multiple parameters has only been demonstrated with optical systems\cite{SU11,Joint_measurement,dense_code,dense_metrology}. Here we report the estimation of multiple simultaneous displacements on the trapped ion system with motional states squeezed along multiple axes. A related recent study\cite{You} reports the simultaneous estimation of multiple phases beyond the standard quantum limit, in a spinor atomic Bose-Einstein condensate (BEC) squeezed along multiple axes.

In contrast to optical systems, we employ reservoir engineering\cite{reservoir} to generate squeezed motional states along multiple axes. Reservoir engineering is a deterministic and efficient approach that allows for the stable generation of squeezed states by harnessing engineered dissipation. This approach stands out from other techniques as it converts a source of noise, which would typically be undesirable, into a valuable resource. By doing so, reservoir engineering allows for the transformation of any input state into the desired state. Reservoir engineering has been proven successful in generating and stabilizing quantum superposition states of single oscillator modes as well as entangled states of two mechanical oscillators\cite{massive, micromechanical,Kienzler, GKP,GKP_Error_correction,GKP_gate}. Theoretical proposals also exists for the generation of two-mode squeezed states using reservoir engineering with a three-mode system\cite{optomechanics, Entanglement_Optomechanical}. Recently, we notice that TMSS can be also prepared in a trapped ion system with a laser-free method using additional radio potentials, which is utilized to demonstrate the SU(1,1) interferometers \cite{Allcock2023}.

In our experimental setup, we utilize the quantized secular motional modes of a trapped ion system coupled to the internal spin of trapped ions through coherent laser fields\cite{RevModPhys.75.281}. The spin serves as a reservoir for the stable production of motional states\cite{Kienzler, GKP,GKP_Error_correction,GKP_gate} and as a coherent channel for high precision motional control\cite{HOM, phononic_network, Multimode, NOON}. Specifically, we employ reservoir engineering to generate a stable two-mode squeezed state along two secular motional modes of a trapped ion. We then utilize this generated state to accurately estimate two collective displacements along the squeezed axes, surpassing the standard quantum limit for both parameters. Our experimental results demonstrate a significant enhancement in precision, reaching up to 6.9(3)~dB and 7.0(3)~dB for the estimation of displacements along the two squeezed axes. These improvements are achieved with a fidelity of the target squeezed state estimated to be over 87(6)\%. Additionally, we confirm the entanglement between the two modes by measuring the Einstein-Podolsky-Rosen (EPR) type variance (also known as the Duan-Criterion)\cite{Duan-Criterion}, which yields a value of 0.20(2), deep below the classical threshold of 1. Importantly, it is not feasible to achieve such simultaneous enhancements in the parameter estimation of collective displacements for a non-entangled product state of squeezed modes. These demonstrations have paved the way for harnessing reservoir engineering to create stable multi-mode squeezed states, unlocking a wealth of possibilities for utilizing squeezed states for multi-parameter quantum metrology. This brings us one step closer to realizing the practicality required for real-world applications. 

The experimental procedure is depicted in Figure~1. We consider two vibrational modes of a trapped ion. The annihilation and creation operators of these modes are denoted as $a_k$ and $a^{\dagger}_k$, and the energy-eigenstates are denoted as $|n\rangle_k$, with $k=1,2$ and $n=0,1,2...$. We start from an uncoupled thermal state after the basic Doppler and electromagnetically-induced-transparency (EIT) cooling\cite{EIT}. We then use laser fields to couple both modes to the same spin, enabling a collective reservoir engineering with controlled dissipation on the spin. The process, which will be described in detail later, allows us to effectively pump and stabilize both modes into a collective two-mode squeezed state. Mathematically, this state can be represented as $|\psi_{\rm TMSS}\rangle=\hat{S}(\xi) |0_1,0_2\rangle$, where $\hat{S}(\xi) = \exp[\xi a_1a_2 - \xi^*a_1^{\dagger}a_2^{\dagger}]$ is the two mode squeezing operator with squeezing parameter $\xi = r e^{i\phi}$. Without loss of generality, we set $\phi=0$ throughout this work. One of the most notable characteristics of $|\psi_{\rm TMSS}\rangle$ is that the variances along two orthometric axes, $\hat{X}_{+} = \frac{1}{\sqrt{2}}(\hat{X}_{1}+\hat{X}_{2})$ and $\hat{P}_{-} = \frac{1}{\sqrt{2}}(\hat{P}_{1}-\hat{P}_{2})$, are simultaneously squeezed by $e^{-2r}$. Here $\hat{X}_{1,2} = \frac{1}{\sqrt{2}}(a_{1,2}+a_{1,2}^{\dagger})$ and $\hat{P}_{1,2} = -\frac{i}{\sqrt{2}}(a_{1,2}-a_{1,2}^{\dagger})$. We then apply simultaneous displacements along the respective squeezed axes. These displacements are implemented using the unitary operator $U = \exp\{-i\Omega_+\hat{P}_+ t\}\exp\{-i\Omega_-\hat{X}_- t\}$, where $\hat{P}_{+} = \frac{1}{\sqrt{2}}(\hat{P}_{1}+\hat{P}_{2})$ and $\hat{X}_{-} = \frac{1}{\sqrt{2}}(\hat{X}_{1}-\hat{X}_{2})$ are the operators conjugate to the squeezed axes. Due to the simultaneous squeezing, the precision for the estimation of both displacement parameters, $\Omega_{\pm}$, are improved exponentially with respect to the squeezing parameter, $r$. After an interaction time $t$, the collective two-mode squeezed state exhibits specific expectation values and variances that are relevant for the estimation of the displacement parameters $\Omega_+$ and $\Omega_-$. The expectation value of the operator $ \hat{P}_-(\hat{X}_+)$ is given by $ \langle \hat{P}_-\rangle = -\Omega_- t~ (\langle \hat{X}_+\rangle = \Omega_+ t)$, 
thus the average displacement along the squeezed axis $ P_- (X_+)$ is directly proportional to the parameter $\Omega_-(\Omega_+)$. 
The variances of the operators $\hat{P}_-$ and $\hat{X}_+$ are given by $(\Delta \hat{P}_-)^2 = (\Delta \hat{X}_+)^2 = \frac{1}{2} e^{-2r}$, where $(\Delta \hat{A})^2=\langle\hat{A}^2\rangle-\langle\hat{A}\rangle^2$. They decrease exponentially with the squeezing parameter, resulting in reduced fluctuations along the respective directions. The enhancement factor of $e^{-2r}$ is consistent with the level of squeezing achieved simultaneously.
These expectation values and variances enable precise estimation of the parameters $\Omega_\pm$, as $\Omega_+ = \langle \hat{X}_+\rangle/t$ and $
\Omega_- = -\langle \hat{P}_-\rangle/t$, whose variances are given by

\begin{eqnarray}
 \delta\Omega_{\pm}^2 = \frac{1}{2}e^{-2r} t^{-2}.
\end{eqnarray}
It achieves exponential enhancement with respect to the squeezing parameter and a Heisenberg scaling with respect to $t$. 
This represents a significant improvement over the classical limits. Since $P_+$ and $X_-$ commute with each other, the two observables can be measured simultaneously without any back-action. This measurement scheme is optimal for the simultaneous estimation of $\Omega_{\pm}$, evident by the saturation of the quantum Cramér-Rao bound as shown in the  supplementary materials\cite{supp}. 

\begin{figure}[!ht]
    \centering
    \includegraphics[width=\columnwidth]{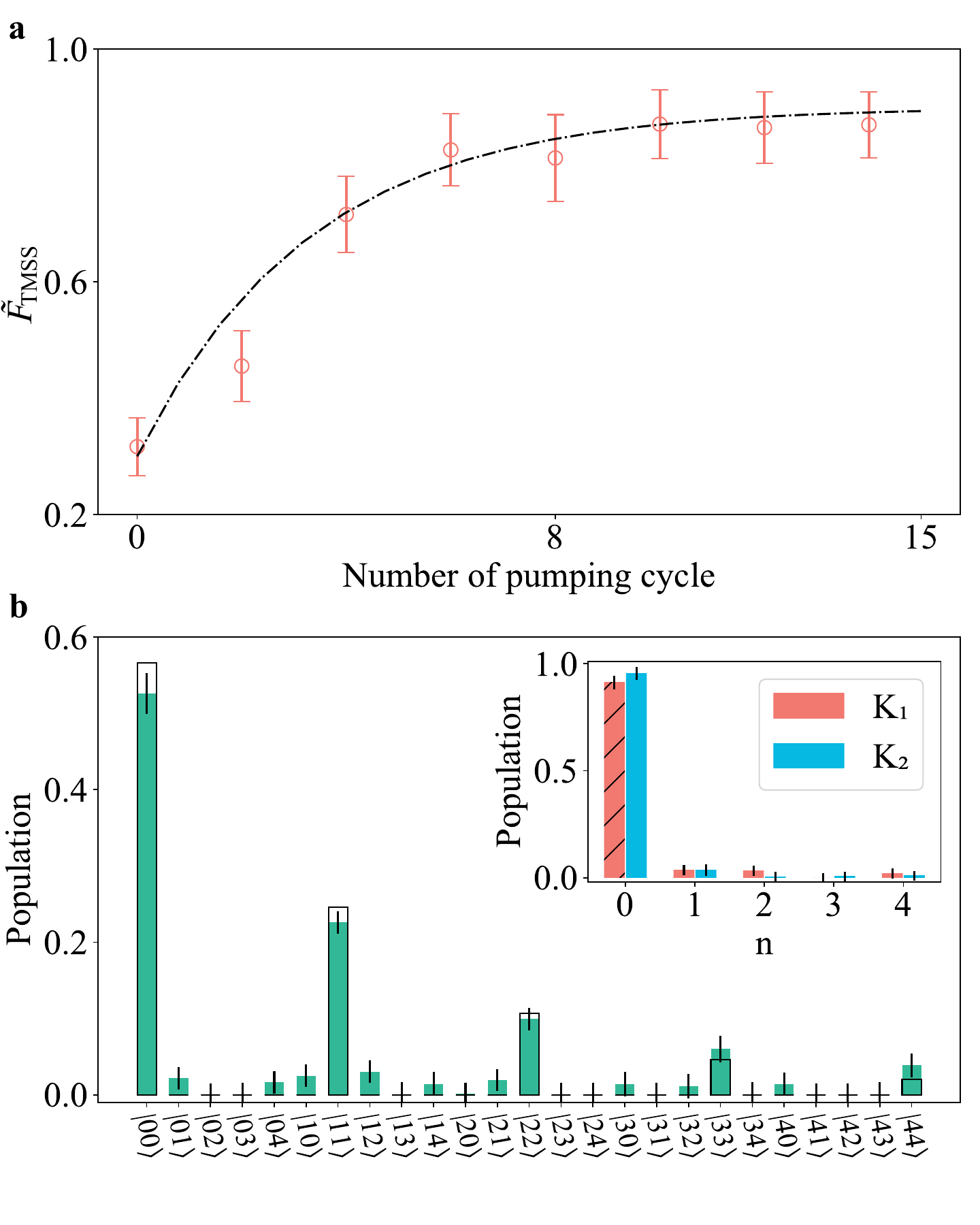}
    \caption{\textbf{Experimental results of a stabilized two-mode squeezed state with squeezed parameter $r = 0.79$.} \textbf{a,} We apply the engineered Hamiltonian and optical pumping by sequence. The experimental data demonstrates the attainment of a steady state after approximately 10 dissipative process cycles, with $\tilde{F}_{\rm TMSS}$ denoting the lower bound of the fidelity for the desired TMSS. The error bars are deduced from the standard error of the fitted parameters. \textbf{b,}  The green bar charts show the experimental fitted population of Fock states $|i,j\rangle$ with $i,j\in[0,4]$, and the black frames show the theoretical population. In the inset figure, the bar charts show the population in the Bogoliubov-transformed basis, represented by $K_{1,2}$. The original data and analysis can be found in supplementary materials\cite{supp}.}
\end{figure}

To generate and stabilize the two-mode squeezed state through reservoir engineering, we employ laser interactions that couple the motional modes to a spin as a zero-temperature bath, and use the spin to effectively cool the motion to the desired stable state. 
The Hamiltonian governing these interactions is given by $\hat{H_i}=\hbar(\Omega_i\hat{K_i}\sigma^++\Omega_i^*\hat{K_i}^\dagger\sigma^-)$, here $\sigma^{\pm}$ correspond to the spin raising and lowering operators, $\hat{K_i}$ are the Bogoliubov-transformed operators for the modes\cite{Bogoliubov} with $\hat{K}_i = \hat{S}(r)a_i\hat{S}(r)^{\dagger}$, where $i=\{1,2\}$ (the same below). 
This Hamiltonian reduces the energy of the motional states in the two-mode squeezed eigen-basis, while flopping the spin from $|\!\!\downarrow \rangle$ to $|\!\!\uparrow \rangle$,
then we add an optical pumping process for re-initialization to $|\!\!\downarrow \rangle$ as shown in Fig. 1d. Briefly, a coherent drive collocating with an optical pump can reduce the entropy. With this dissipation cycle, we can produce a zero-energy state $\hat{S}(r)|0,0\rangle$. Explicitly, a pair of independent laser beams are employed, 
with the ratio of their strength quantified by $\tanh(r)$, is used to control the squeezing parameter $r$ while the associated Bogoliubov-transformation is given by
\begin{equation}
    \hat{K}_i = \hat{S}(r)a_i\hat{S}(r)^{\dagger} = a_i \cosh(r) + a_{j\neq i}^{\dagger}\sinh(r).
\end{equation}
The coupling strength between the modes and the spin, denoted as $\Omega_i$, is determined by the absolute intensities of the laser fields. The two-mode squeezed state is then the collective ground state of $\hat{K}_{i=1,2}$ with $\hat{K}_{i=1,2}|\psi_{\rm TMSS}\rangle=0$ , which can be obtained by cooling the motional state. With alternating application of the above process between the two modes for a number of repetitions, as depicted in Figure~2b, the collective ground state, $|\psi_{\rm TMSS}\rangle$, can be ideally reached.

In the experiment, we use a $^{40}\rm{Ca}^+$ trapped in a cryogenic linear~Paul trap\cite{RevModPhys.75.281}. The trap frequencies used in this work are $\omega_1 =2\pi \times 1.12$ MHz and $\omega_2 =2\pi \times 0.90$ MHz. 
Before each experiment, the motional modes are initialized by the cooling processes mentioned above, reducing the mean phonon number of both modes to a thermal state of average excitation of less than 0.2 phonon\cite{supp}. For the spin degree of freedom, we employ the electronic states $|\!\!\downarrow\rangle \equiv |L=0, J=1/2, M_J = + 1/2\rangle$ and $|\!\!\uparrow\rangle \equiv |L=2, J=5/2, M_J = + 1/2\rangle$, with a resonant frequency difference denoted as $\omega_0$. To achieve optical pumping from all other states to $|\!\!\downarrow\rangle$, we utilize a combination of a $\sigma^{+}$ polarized 397~nm laser and linearly polarized 866~nm and 854~nm lasers. This process is denoted as $L_{|\uparrow\rangle\rightarrow|\downarrow\rangle}$ \cite{supp}. To achieve spin-motion coupling, we utilize 729~nm lasers with frequencies of $\omega_0\pm\omega_{i=1,2}$, driving transitions between the states $|\!\!\downarrow,n_i\rangle$ and $|\!\!\uparrow,n_i\pm1\rangle$. These transitions are commonly referred to as sideband drives. We have independent control over the frequency, phase, and amplitude of all four components of the laser fields. By simultaneously applying these laser fields in the desired combinations, we can achieve the desired coupling described by $\hat{H}_{1,2}$. We apply the sequence $\hat{H}_1-L_{|\uparrow\rangle\rightarrow|\downarrow\rangle}-\hat{H}_2-L_{|\uparrow\rangle\rightarrow|\downarrow\rangle}$ repeatedly for a total of $N$ times. The applied strengths of $H_1$ and $H_2$ are equal and denoted as $\Omega_1 = \Omega_2 = 2\pi\times 6.8(1)~\rm{kHz}$. Each $\hat{H}_1$ or $\hat{H}_2$ is applied for a duration of $55~\rm{\mu s}$. This sequence generates the desired state $|\psi_{\rm TMSS}\rangle$, which is quantified through the detection process described below. In Fig.~2a, we present the process of producing $|\psi_{\rm TMSS}\rangle$ for a specific value of $r=0.79$, along with the corresponding value of $N$. To evaluate the fidelity, we analyze the output state $\rho$ in terms of the motional population $P(n^K_{i})$. Here, $|n^K_{i}\rangle=\hat{S}(r)|n_i\rangle$ represents the Bogoliubov-transformed Fock state for mode $i=1,2$, and $P(n^K_{i})$ is calculated as $Tr(\langle n^K_i|\rho|n^K_i\rangle)$. The lower bound of the fidelity $\tilde{F}_{\rm TMSS}$ is then obtained as $\tilde{F}_{\rm TMSS}=P(0^K_1)P(0^K_2)$. We find that the lower bound $\tilde{F}_{\rm TMSS}$ increases with the number of cycles $N$, reaching a steady-state after approximately 10 cycles of pumping, with a fidelity measured up to $87(6)\%$.

The measurement is achieved by first applying a time-varying drive with the analysis Hamiltonian $\hat{H}^{+}_{1,2} = \hbar \Omega(\hat{K}_{1,2}\sigma^{-} + \hat{K}_{1,2}^{\dagger} \sigma^{+})$ followed by spin fluorescence detection. The resulting time-dependent curve of the spin population exhibits multiple sinusoidal oscillations corresponding to the population of $|n^K_{i=1,2}\rangle$, as shown in the inset of Fig.~2b, where we can observe the majority population concentrates in the Bogoliubov-transformed ground state, indicating successful generation of the target state. To extract the joint motional population in the Fock basis, we employ curve-fitting techniques with a model that describes the behavior of the spin-motion coupled system. Additionally, we develop a technique to obtain the joint motional population $P(n_1,n_2)$ for the state $|n_1, n_2\rangle$ in the original basis (as described in the supplementary materials\cite{supp}). To achieve this, we introduce an auxiliary level $|\rm{AUX}\rangle \equiv |L=2, J=5/2, M_J = + 5/2\rangle$. By sequentially performing blue sideband transitions, namely $|\!\!\downarrow,n_1\rangle \leftrightarrow |\!\!\uparrow,n_1+1\rangle$ for a duration $t_1$, and $|\!\!\downarrow,n_2\rangle \leftrightarrow |\rm{AUX},n_2+1\rangle$ for a duration $t_2$, we can express the population of the state $|\!\!\downarrow\rangle$ as a function of the durations $t_1$, $t_2$, and $P(n_1,n_2)$. The joint motional population $P(n_1,n_2)$ can then be extracted by performing a two-dimensional curve-fitting. The fitting results for the targeted TMSS with a squeezed parameter $r=0.79$ are displayed in Fig.~2b. Since $|\psi_{\rm TMSS}\rangle =\sum_{i=0,1,2...} c_i|i_1, i_2\rangle$ with $i_1=i_2$ in the original Fock basis, we experimentally observe the majority of populations in correlated pairs such as $|0,0\rangle$, $|1,1\rangle$, $|2,2\rangle$ and $|3,3\rangle$, which aligns well with our theoretical expectations.

\begin{figure}[!t]
    \centering
    \includegraphics[width=\columnwidth]{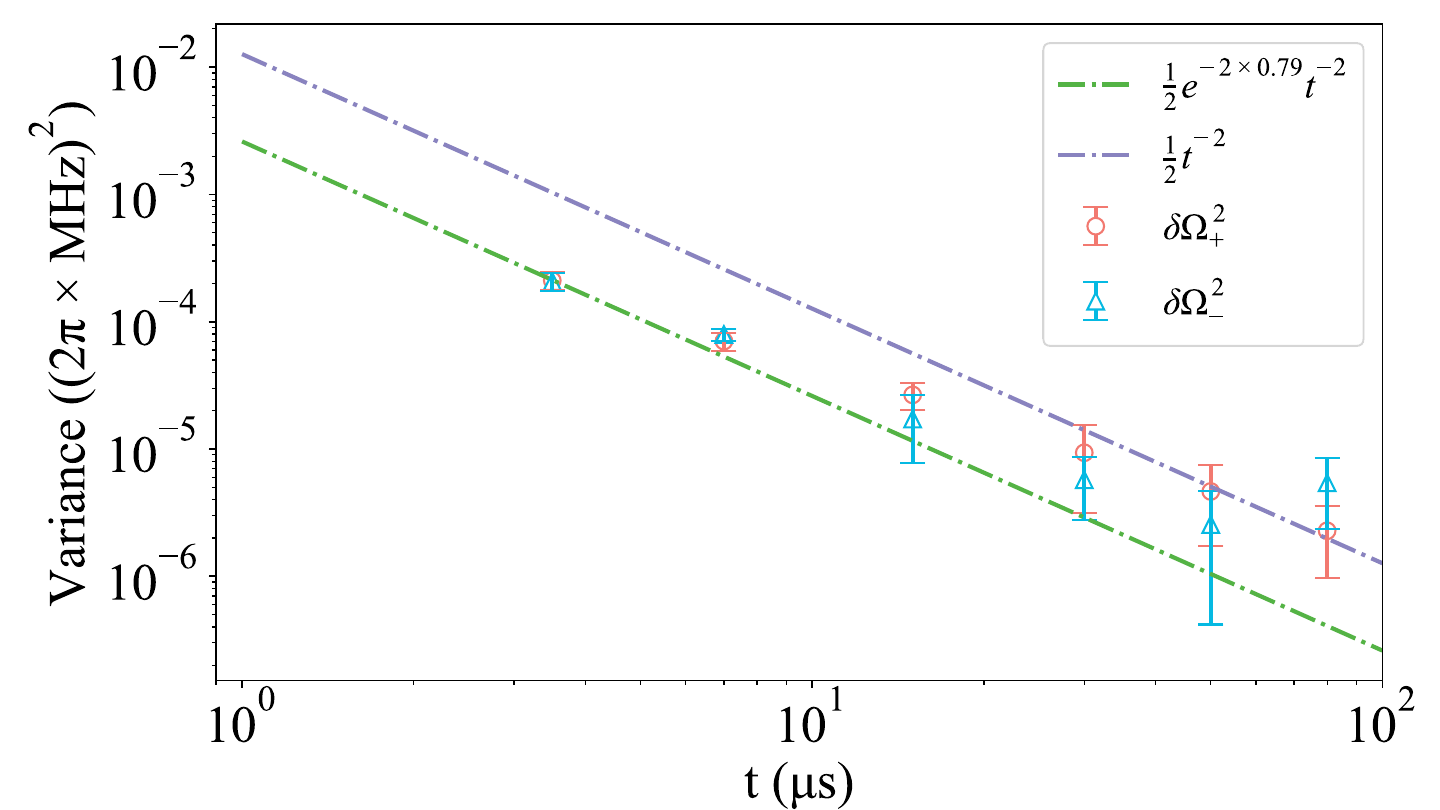}
    \caption{\textbf{Experimental results of the variance of $\Omega_+$  and $\Omega_-$ in the simultaneous estimation.} In our experiment, we select a squeezed parameter $r=0.79$. The blue triangle and red circle data points show the experimental result for the two parameters $\Omega_\pm$. The green dashed line shows the theoretical best precision result $\delta\Omega_+ ^2 = \delta\Omega_-^2 = \frac{1}{2}e^{-2\times 0.79} t^{-2}$, which corresponding to Heisenberg limit.  The purple dashed line shows the classical limits, and our experiment demonstrates up to 6.9(3)~dB and 7.0(3)~dB enhancement over the classical limits for the two parameters $\Omega_\pm$ respectively.  Each measurement is obtained from 200 trials.}
\end{figure}

In order to demonstrate multi-parameter metrology with the state $|\psi_{\rm TMSS}\rangle$, we employ a combination of laser sideband drives on the same mode, which results in operations governed by the Hamiltonian 
$\hat{H}_+=\hbar\Omega_+\hat{P}_+\sigma_x$  and $\hat{H}_-=\hbar\Omega_-\hat{X}_-\sigma_x$, where $\sigma_x$ is a Pauli spin operator. Considering the estimation of the parameters $\Omega_{\pm}$, here we aim to reduce the variance $\delta\Omega_{\pm}^2$ by utilizing the two-mode squeezed state. After preparing the state $|\psi_{\rm TMSS}\rangle$, we apply a resonant 729~nm laser pulse with a duration of $\pi/2$ that drives the transition between $|\!\!\uparrow\rangle$ and $|\!\!\downarrow\rangle$. This pulse prepares the state $\frac{1}{\sqrt{2}}(|\!\!\uparrow\rangle+|\!\!\downarrow\rangle)$, which is an eigenstate of $\sigma_x$ on the spin. By sequentially applying $H_\pm$ for a duration $t$ respectively, the displacements along the squeezed axes are generated, while the spin state remains unchanged. This creates the desired unitary operator, $U= \exp\{-i\Omega_+\hat{P}_+t\}\exp\{-i\Omega_-\hat{X}_-t\}$, that encodes the parameters on the motional states.

To estimate the parameters and quantify the precision, we measure the variance $\delta\Omega_{\pm}^2$ from
$\langle \hat{X}_+\rangle$, $\langle \hat{P}_-\rangle$, $\langle\hat{X}_+^2\rangle$, and $\langle\hat{P}_-^2\rangle$. We couple the motion to the spin via time-varying spin-dependent displacements $U_p = \exp(-i\Omega_p t\hat{A}\sigma_x)$, with $\hat{A}=\{\hat{X}_+, \hat{P}_-\}$ and $\Omega_p$ the Rabi rate. Consider a spin detection along $\sigma_z$, such operation transforms to a measurement of $\langle U_p^\dagger \sigma_z U_p\rangle =  \langle\sin(\Omega_p t\hat{A})\sigma_y\rangle+\langle\cos(\Omega_p t\hat{A})\sigma_z\rangle$. Thus, by initializing the spin to the eigen-states of  $\sigma_y$ and $\sigma_z$, the  first order and second order time-derivatives for the measuremnt curve give $\langle \hat{A}\rangle$ and $\langle \hat{A^2}\rangle$ respectively, which further leads to $\delta\Omega_{\pm}^2$. In our experimental setup, we generate a TMSS with a squeezing parameter of $r=0.79$ and apply $\Omega_\pm=2\pi\times 5.0~ \rm{kHz}$. In Figure~3 we plot the variance of the measurement result with respect to the duration $t$. We observe that the variance for the estimation of both parameters scale as $t^{-2}$ with an additional exponential reduction of $e^{-2r}$ gained from the squeezing, resulting in a total variance as $\frac{1}{2}e^{-2r}/t^2$, matching with the theoretical predictions. We experimentally observe 6.9(3)~dB and 7.0(3)~dB enhancements for the estimation of two parameters, which equivalently indicates the level of
squeezing of the prepared TMSS along the two axes. At large $t$, the experimental data starts to deviate from the ideal theoretical curve. This can be attributed to dephasing during the preparation of TMSS and subsequent displacements along the $\hat{X}_+$ and $\hat{P}_-$ directions.

\begin{figure}[!t]
    \centering
    \includegraphics[width=\columnwidth]{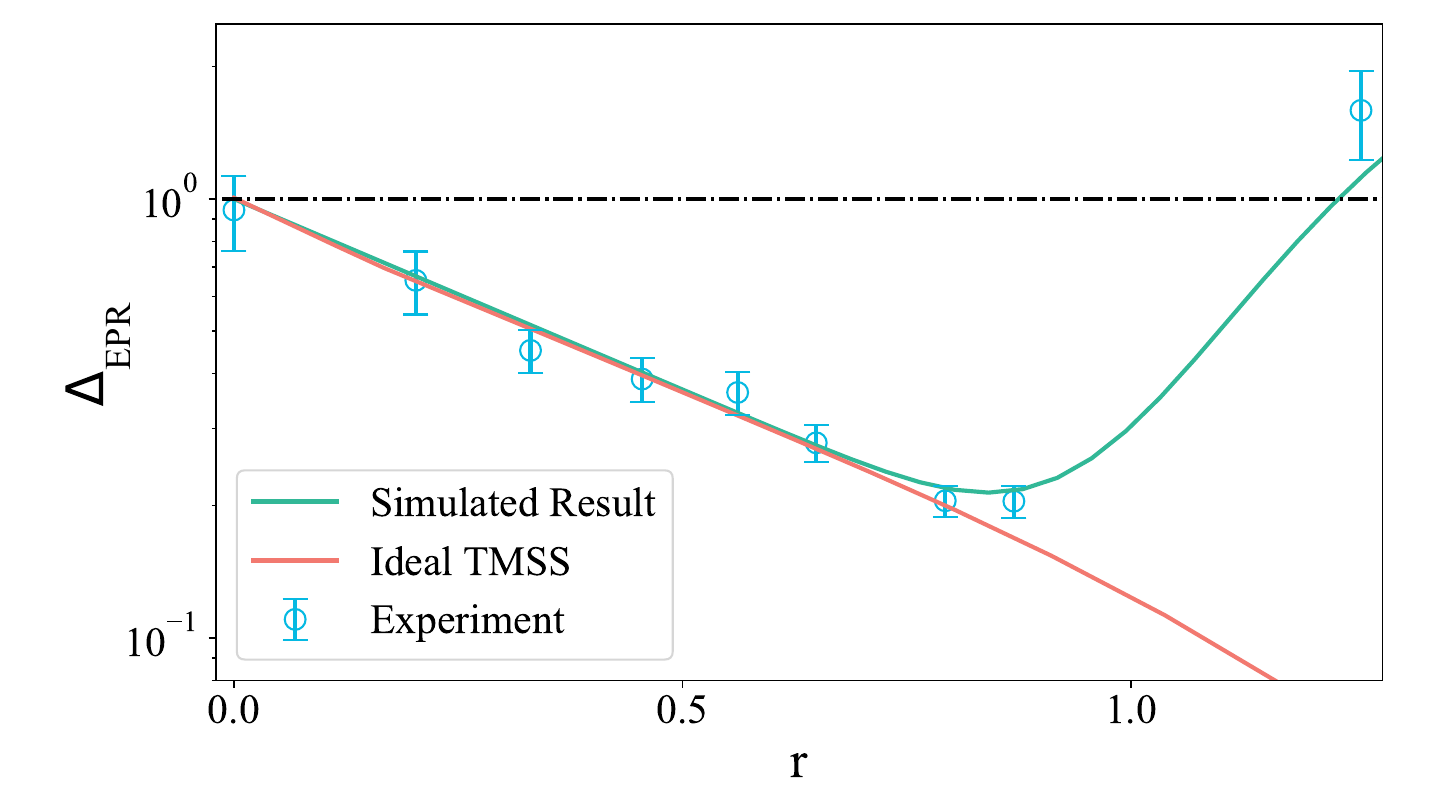}
    \caption{\textbf{Entanglement criterion.} Entanglement expressed via Duan quantity $\Delta_{\rm{EPR}}$ against the squeezed parameter $r$. The red line shows Duan quantity of the ideal TMSS. The green line shows the numerical simulation results of 10 dissipation cycles. The black dashed line shows the Duan bound, below which entanglement can be distinguished under this bound.}
\end{figure}

We also investigate the relationship between precision and the entanglement of TMSS, which can be quantified by the variance with respect to Einstein-Podolsky-Rosen (EPR) type operators\cite{Duan-Criterion}, $\Delta_{\rm{EPR}} =  (\Delta \hat{X}_+)^2 + (\Delta \hat{P}_-)^2$. The EPR variance provides a measure of the entanglement and the metrology precision of a two-mode Gaussian state. As quantified by the Duan inequality\cite{Duan-Criterion}, a two-mode Gaussian state is entangled if $\Delta_{\rm{EPR}} < 1$, where a lower value of the EPR variance indicates stronger entanglement and thus higher measurement precision. In our experiment, we prepared a TMSS state with varied squeezing parameter $r$, and we measured the values of corresponding $\Delta_{\rm{EPR}}$ as depicted in Fig.~4. We applied 10 dissipation cycles and systematically varied the squeezed parameter $r$ to examine its impact on the entanglement of the two motional modes. As shown in Figure~4,  $\Delta_{\rm{EPR}}$ saturates near $r=0.8$, matched with our model considering experimental imperfections\cite{supp}. In particular, with $r=0.79$, we obtain $\Delta_{\rm{EPR}} = 0.20(2)$ close to the ideal TMSS value of $\Delta_{\rm{EPR}} = 0.205$, corresponding to a strong entanglement and a precision enhancement of 7.0~dB. For larger squeezing values ($r > 1.0$), more dissipation cycles are needed, as detailed in supplementary materials\cite{supp}. Deviations from the ideal for the case of larger squeezing values are due to the technically limited amount of pulses applied leading to a state that has not yet reached the steady state. These deviations could be alleviated by adding more dissipation cycles, as predicted in the theoretical analysis.

In summary, our experiment demonstrates a successful utilization of states squeezed along multiple axes to achieve simultaneous estimation of multiple parameters with exponential quantum enhancement. We employ a scheme based on reservoir engineering to generate entanglement between quantum harmonic oscillators in a stable manner. Using a two-mode squeezed state (TMSS) characterized by a squeezing parameter of $r=0.79$, we experimentally observe a precision improvement of 6.9(3)~dB and 7.0(3)~dB in the simultaneous estimation of two parameters, respectively. To validate the generation of TMSS, we measure the fidelity, which reaches up to $87(6)\%$, and directly observe the population distribution in the Fock basis, confirming our expectations. We also quantify the entanglement of TMSS by measuring the EPR variance, with a value of $\Delta_{\rm{EPR}} = 0.20(2)$, deep below the criterion for entanglement($\Delta_{\rm{EPR}}<1$). Our setup can be readily upgraded to incorporate single-shot readout of the multiple levels in a trapped ion\cite{Ringbauer2022}, to facilitate the measurements of multi-parameters. Taking advantage of our precise control over the coupling between multiple motional modes and the internal state of ions, we can scale our system to accommodate a larger number of ions and motional modes by employing lasers that address individual ions. This scalability allows us to expand our toolbox and generate squeezed states with even more modes, enabling the simultaneous estimation of multiple parameters. To demonstrate the scalability, we have experimentally prepared a three-mode squeezed state with a squeezing parameter of $r=0.5$ and fidelity of 88(7)\%, as detailed in the supplementary materials\cite{supp}. This state can be utilized for the estimation of three parameters, highlighting the versatility of our approach. 
With the capability to extend the number of motional modes to over one hundred\cite{200}, our approach paves the way for the exploration of multi-parameter quantum-enhanced metrology using states squeezed along multiple axes. This approach also holds vast potential for applications in quantum simulation, facilitating the study of many-body correlated states, and in the field of quantum information processing with continuous variables. Since our approach demonstrates a step-wise quantum control with dissipation, further generalization may be possible for a new path of quantum metrology with open quantum system, when considering a combination of recent advances with quantum circuit based metrology\cite{Marciniak2022,Conlon2023} and the technique of complex state engineering with dissipation\cite{Verstraete2009}. Importantly, this approach is not confined to a specific experimental platform as it can be implemented in a wide range of systems including nano-mechanics and superconducting circuits.\\\\

 We thank J. J. Bollinger and L. You for helpful comments, and X. Rong, F. Shi, Y. Wang, and T. Xie for apparatus supports. The USTC team acknowledges support from the National Natural Science Foundation of China (grant number 92165206, 11974330), Innovation Program for Quantum Science and Technology (Grant No. 2021ZD0301603), Anhui Initiative in Quantum Information Technologies (Grant No. AHY050000), the USTC start-up funding, and the Fundamental Research Funds for the Central Universities. H.Y. acknowledges partial support from the Research Grants Council of Hong Kong with Grant No. 14307420, 14308019,14309022.
\\




\section{Supplemental Material}

\subsection{Theoretical Derivation of TMSS Quantum Metrology}
The two-mode squeezed state is defined by $\hat{S}(\xi) |0,0 \rangle$, here 
\begin{equation}
  \hat{S}(\xi) = \exp\{\xi a_1 a_2 - \xi^{*} a_1^{\dagger} a_2^{\dagger}\},
\end{equation}
is the two-mode squeeze operator with a squeezing parameter $\xi = r e^{i \phi}$.
We can choose $\phi = 0$ without loss of generality.

Let the initial state $|\psi_0\rangle = \hat{S}(r)|0,0\rangle$ go through the evolution process
\begin{equation}
  U = \exp\{-i \Omega_- \hat{X}_{-}  t\} \exp\{-i \Omega_+ \hat{P}_{+} t\}.
\end{equation}
Then the evolved state is $|\psi_t\rangle = U |\psi_0\rangle$, which encoded with the unknown parameters $\Omega_-$ and $\Omega_+$.
Here
\begin{equation}
    \hat{X}_{\pm} = (\hat{X}_1 \pm \hat{X}_2)/\sqrt{2}, \quad
    \hat{P}_{\pm} = (\hat{P}_1 \pm \hat{P}_2)/\sqrt{2}, 
\end{equation}
and for $k=1,2$,
\begin{equation}
    \hat{X}_{k} = (a_{k}+a_{k}^{\dagger})/\sqrt{2}, \quad
    \hat{P}_{k} = -i (a_{k}-a_{k}^{\dagger})/\sqrt{2}.
\end{equation}
Using the fact that $[\hat{X}_k, \hat{P}_k] = i$, where we set $\hbar=1$, we have
\begin{equation}
  [\hat{X}_+, \hat{P}_-] = 0, \quad [\hat{X}_-, \hat{P}_+] = 0.
\end{equation}

The generator of $U$ for an unknown parameter $x$ is defined by $H_x = i U^{\dagger}\left(\partial_{x} U\right)$, thus the generators w.r.t parameters $\Omega_-$, $\Omega_+$ can be obtained as
\begin{equation}
    H_{\Omega_-} = \hat{X}_{-} t, \quad
    H_{\Omega_+} = \hat{P}_{+} t,
\end{equation}
respectively. Then the elements of the quantum Fisher information matrix can be calculated from
\begin{equation}
    \left[F_Q\right]_{m n} = 2\left\langle\psi_0\left|\{H_{m}, H_{n}\}\right| \psi_0\right\rangle-4 \left\langle\psi_0\left|H_{m}\right| \psi_0\right\rangle \left\langle\psi_0\left|H_{n}\right| \psi_0\right\rangle, 
\end{equation}
with $m,n \in \{\Omega_-, \Omega_+\}$.
By using the properties that

\begin{equation}
  \begin{aligned}
    &\hat{S}(r)^{\dagger} a_1 \hat{S}(r) = a_1 \cosh r - a_2^{\dagger} \sinh r \\
    &\hat{S}(r)^{\dagger} a_2 \hat{S}(r) = a_2 \cosh r - a_1^{\dagger} \sinh r
  \end{aligned}
\end{equation}

we can show that
\begin{eqnarray}
\langle 0,0| \hat{S}(r)^{\dagger} \hat{X}_{-} \hat{S}(r) |0,0\rangle =0, \\
\langle 0,0| \hat{S}(r)^{\dagger} \hat{P}_{+} \hat{S}(r) |0,0\rangle = 0,
\end{eqnarray}
and
\begin{eqnarray}
\langle 0,0| \hat{S}(r)^{\dagger} \hat{X}_{-}^2 \hat{S}(r) |0,0\rangle = \frac{1}{2} e^{2r}, \\
\langle 0,0| \hat{S}(r)^{\dagger} \hat{P}_{+}^2 \hat{S}(r) |0,0\rangle = \frac{1}{2} e^{2r}, \\
\langle 0,0| \hat{S}(r)^{\dagger} \left\{\hat{X}_{-}, \hat{P}_{+} \right\} \hat{S}(r) |0,0\rangle =0. 
\end{eqnarray}

The quantum Fisher matrix can thus be obtained as
\begin{equation}
  F_Q = \left(\begin{array}{cc}
    2 e^{2 r} t^2 & 0 \\
    0 & 2 e^{2 r} t^2
  \end{array}\right),
\end{equation}

from which we can quantify the precision of simultaneous measurement of parameters $\Omega_-$ and $\Omega_+$ as
$Tr(F_Q^{-1}) = e^{-2r} t^{-2}.$
  
We note that the generators for two parameters commute, thus there exist some measurements to achieve this precision. 
We now present the optimal observables for estimating parameter $\Omega_-$ and $\Omega_+$ are $\hat{P}_{-}$ and $\hat{X}_{+}$, respectively.

The two observables $\hat{X}_{+}$ and $\hat{P}_{-}$ are commuting, thus that can be jointly measured.
Consider the joint measurement of $\hat{X}_{+}$ and $\hat{P}_{-}$, denote
$|\chi, \eta\rangle$ as the simultaneous eigenstates of the commuting observables $\hat{X}_{+}$ and $\hat{P}_{-}$ with eigenvalues $\chi$ and $\eta$ respectively, which is given by\cite{helstrom1973cramer}
\begin{equation}
  \begin{aligned}
    \ket{\chi, \eta } &= \pi^{-1/2} \int e^{2 i \eta x} \ket{x}_1 \otimes \ket{\chi - x}_2 dx \\
    &=\pi^{-1/2} \int e^{2 i (\eta - p)\chi} \ket{p}_1 \otimes \ket{p - \eta}_2 dp
  \end{aligned}
\end{equation}
where $|x\rangle_1$ and $|\chi -x\rangle_2$ are the eigenstates of $\hat{X}_1$ and $\hat{X}_2$ with eigenvalues $\sqrt{2} x$ and $\sqrt{2}(\chi-x)$, respectively.
$|p\rangle_1$ and $|p -\eta\rangle_2$ are the eigenstates of $\hat{P}_1$ and $\hat{P}_2$ with eigenvalues $\sqrt{2}p$ and $\sqrt{2}(p - \eta)$, respectively.
We can check that

\begin{equation}
  \begin{aligned}
    \hat{X}_{+} \ket{\chi, \eta } &= \pi^{-1/2} \int e^{2 i \eta x} \left(x +(\chi-x)\right) \ket{x}_1 \otimes \ket{\chi - x}_2 dx \\ 
    &= \chi \ket{\chi, \eta } ,\\
    \hat{P}_{-} \ket{\chi, \eta } &= \pi^{-1/2} \int e^{2 i (\eta - p)\chi} \left(p-(p-\eta)\right) \ket{p}_1 \otimes \ket{p - \eta}_2 dp \\
    &= \eta \ket{\chi, \eta }. \\
  \end{aligned}
\end{equation}
$|\chi, \eta \rangle$ are normalized so that $\langle \chi^{\prime}, \eta^{\prime}| {\chi,\eta} \rangle = \delta(\chi-\chi^{\prime})\delta (\eta - \eta^{\prime})$. 
Moreover, $U$ can be rewritten as
\begin{equation}
  U = \int \! \! \! \int  e^{-i t(\omega \Omega_- + \tau \Omega_+) } |\omega,\tau\rangle \langle \omega, \tau| d\omega d\tau,
\end{equation}
here $|\omega,\tau\rangle$ are the simultaneous eigenstates of commuting observables $\hat{X}_{-}$ and $\hat{P}_{+}$ with eigenvalues $\omega$ and $\tau$ respectively, which is given by

\begin{equation}
  \begin{aligned}
    \ket{\omega, \tau } &= \pi^{-1/2} \int e^{2 i \tau x} \ket{x}_1 \otimes \ket{x-\omega}_2 dx \\
    &=\pi^{-1/2} \int e^{2 i (\tau - p)\omega} \ket{p}_1 \otimes \ket{\tau-p}_2 dp.
  \end{aligned}
\end{equation}
$|\omega, \tau\rangle$ are normalized so that $\langle \omega^{\prime}, \tau^{\prime}| \omega, \tau \rangle = \delta (\omega^{\prime}-\omega) \delta (\tau^{\prime}-\tau)$.
Similarly, it is easy to verify that $\hat{X}_{-} |\omega, \tau \rangle = \omega |\omega, \tau \rangle$ and $\hat{P}_{+} |\omega, \tau \rangle = \tau |\omega, \tau \rangle$.

We then calculate the joint probability density function of the measurement results from which the following identities will be used,
\begin{equation}
  \begin{aligned}
    &\langle \chi, \eta |{\omega, \tau} \rangle \\ 
    =& \pi^{-1} \iint e^{-2i \eta x} e^{2i\tau x^{\prime}} \langle{x}|{x^{\prime}}\rangle \langle{\chi -x}|{x^{\prime}- \omega}\rangle dx dx^{\prime} \\
    =&\pi^{-1} \iint e^{-2i \eta x} e^{2i\tau x^{\prime}} \delta(x - x^{\prime})  \delta(\chi -x - x^{\prime} + \omega) dx dx^{\prime} \\
    =&\pi^{-1} \int e^{-2i \eta x} e^{2i\tau x}  \delta(\chi+ \omega- 2 x) dx \\
    =& (2\pi)^{-1} e^{-i(\eta -\tau)(\omega + \chi)}.
  \end{aligned}
\end{equation}
The two-mode squeezed state in the coordinate representation is given by the wave function\cite{hong1989simple}
\begin{equation}
\begin{aligned}
    &\langle x_1, x_2| \hat{S}(r) |0,0\rangle \\
    =& \left(\frac{2}{\pi}\right)^{1/2} \exp[-(x_1^2+x_2^2)\cosh 2 r - 2 x_1 x_2 \sinh 2r],
      \end{aligned}
\end{equation}
from which we can get
\begin{eqnarray}
    \langle \omega, \tau| \hat{S}(r) |0,0\rangle
    = \pi^{-1/2} \exp[-r - i \omega \tau -\frac{1}{2} e^{-2r}(\omega^2 + \tau^2)].
\end{eqnarray}
With the above identities, we can get the joint probability density function of the measurement results as
\begin{equation}
    p(\chi, \eta)
     = \frac{1}{\pi}\exp[2r - e^{2r} \left((\chi + \Omega_+ t)^2 + (\eta + \Omega_- t)^2 \right)]
\end{equation}

The expectation value of observables $\hat{P}_{-}$ and $\hat{X}_{+}$ can then be obtained as
\begin{eqnarray}
\langle \hat{P}_{-} \rangle = \int \! \! \! \int \eta p(\chi,\eta) d\chi d\eta = - \Omega_- t \\
\langle \hat{X}_{+} \rangle = \int \! \! \! \int \chi p(\chi,\eta) d\chi d\eta =  \Omega_+ t
\end{eqnarray}
And similarly, 
\begin{eqnarray}
    \langle \hat{P}_{-}^2 \rangle = \int \! \! \! \int \eta^2 p(\chi,\eta) d\chi d\eta =  \frac{1}{2} e^{-2r} + t^2 \Omega_-^2 \\
    \langle \hat{X}_{+}^2 \rangle = \int \! \! \! \int \chi^2 p(\chi,\eta) d\chi d\eta = \frac{1}{2} e^{-2r} + t^2 \Omega_+^2
\end{eqnarray}
Hence, the variance of the observables $\hat{P}_{-}$ and $\hat{X}_{+}$ is
\begin{eqnarray}
  (\Delta \hat{P}_{-})^2 = \langle \hat{P}_{-}^2 \rangle  - \langle \hat{P}_{-} \rangle^2 = \frac{1}{2} e^{-2r}, \\
  (\Delta \hat{X}_{+})^2 = \langle \hat{X}_{+}^2 \rangle  - \langle \hat{X}_{+} \rangle^2 = \frac{1}{2} e^{-2r},
\end{eqnarray}
which gives the variance of estimating parameter $\Omega_-$ and  $\Omega_+$ by error propagation formula,
\begin{eqnarray}
    \delta \Omega_-^2 = \frac{(\Delta \hat{P}_{-})^2}{\left(\partial_{\Omega_-}\langle \hat{P}_{-} \rangle\right)^2} = \frac{1}{2} e^{-2 r} t^{-2},\\
    \delta \Omega_+^2 = \frac{(\Delta \hat{X}_{+})^2}{\left(\partial_{\Omega_-}\langle \hat{X}_{+} \rangle\right)^2} = \frac{1}{2} e^{-2 r} t^{-2}.
\end{eqnarray}
The precision of simultaneous estimating $\Omega_-$ and $\Omega_+$ achieve the Heisenberg scaling and exponential quantum enhancement.

\begin{figure}
    \centering
    \includegraphics[width=\columnwidth]{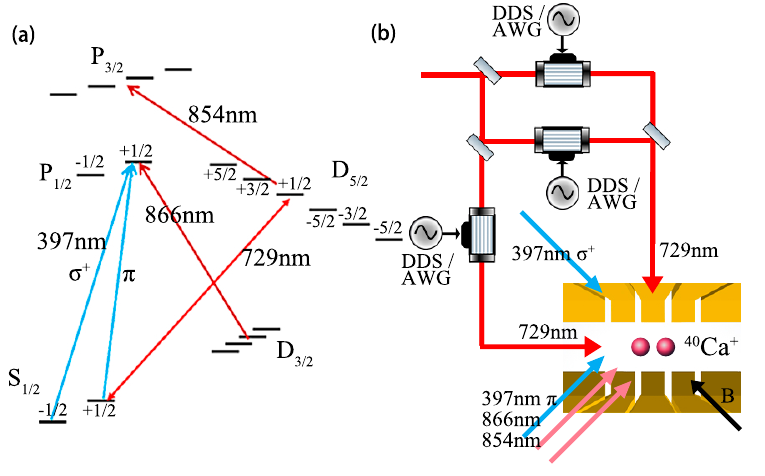}
    \caption{\textbf{Experimental setup.} \textbf{a,} the energy level of $^{40}\rm{Ca}^+$.  \textbf{b,} the optical setup and electronics used in this work.
    \label{fig:S1}
}
\end{figure}

\subsection{Experimental Details}

\begin{figure}
    \centering
    \includegraphics[width=\columnwidth]{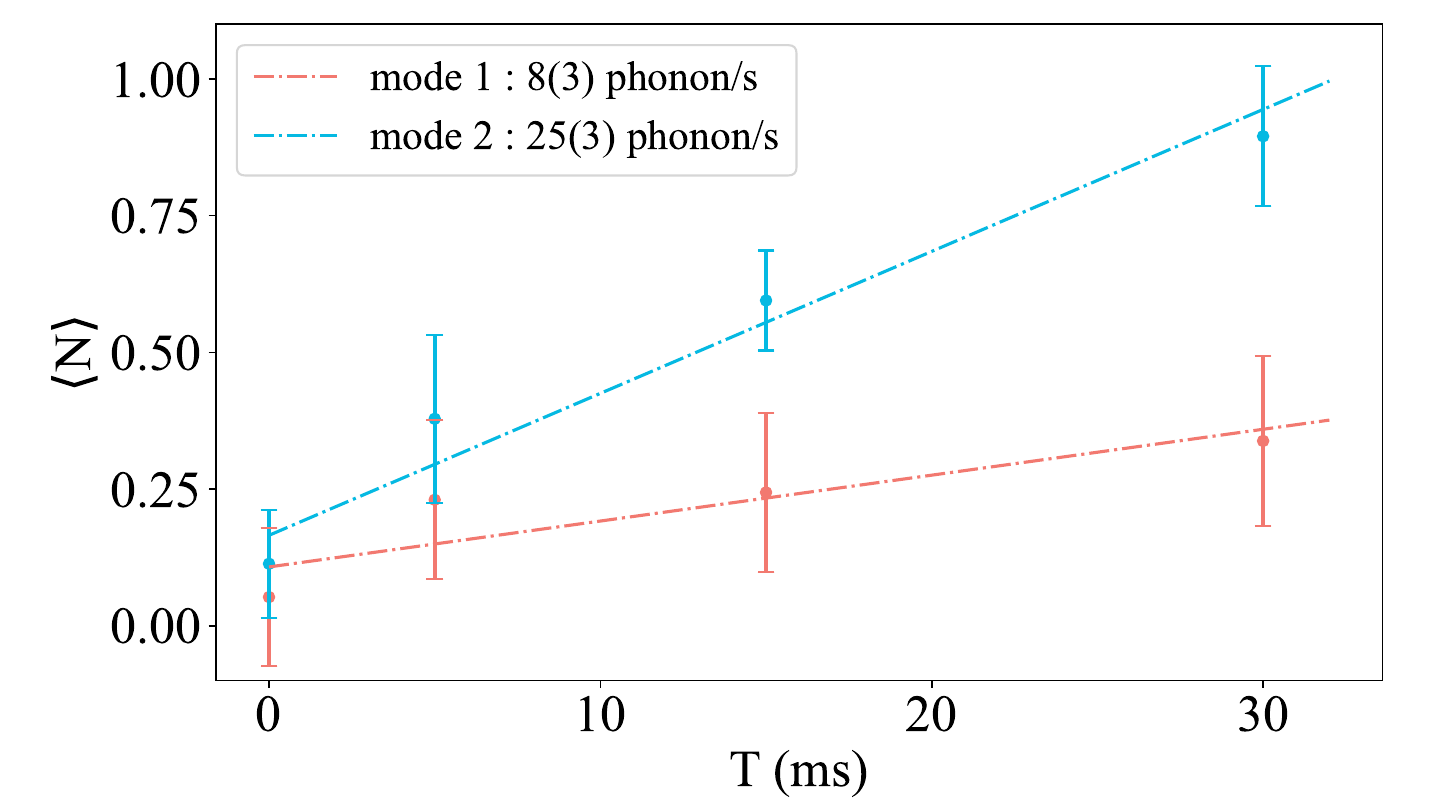}
    \caption{\textbf{Heating rate of two motional mode.} The heating rates for mode 1 and 2 are 8(3) phonon/s and 25(3)phonon/s respectively.}
    \label{fig:S2}
\end{figure}

In this work, we use trapped  $^{40}\rm{Ca}^+$ ions in a linear Paul trap\cite{RevModPhys.75.281} as shown in Figure \ref{fig:S1}. The oscillation modes in the radial directions of trapped ions are denoted as mode 1 and 2, with trap frequencies $\omega_1 =2\pi \times 1.12$ MHz and  $\omega_2 =2\pi \times 0.90$ MHz. The Lamb-Dicke factors of two modes are $\eta_1 = 0.06$, $\eta_2 = 0.07$ respectively, which means that the experiments are well described by the Lamb-Dicke approximation with $\eta\ll1$\cite{RevModPhys.75.281}. We also measure the heating rate of the two mode as shown in Figure~\ref{fig:S2}, which is low enough for our motional state control. Before each experiment, these modes are initialized by Doppler cooling and EIT cooling, as shown in Figure~1.  Doppler cooling is performed via a 866~nm laser resonant with the dipole transitions between $|D_{3/2}\rangle$ and $|P_{1/2}\rangle$, and a 397~nm laser near detuned to the dipole transitions between $|S_{1/2}\rangle$ and $|P_{1/2}\rangle$. A pair of far blue detuned 397~nm laser are used for EIT cooling\cite{EIT}. After EIT cooling, the mean phonon number of these two mode both can be cooled to less than 0.2 phonon. The quadrupole transition between $|S_{1/2}\rangle$ and $|D_{5/2}\rangle$ is used for coherent manipulations drived by a narrow linewidth 729~nm laser. We use two 729~nm lasers along axial and radial directions respectively, which help us control all motional modes along three dimensions. We can perform ground state cooling and coherent control to all six motional modes of two trapped ions. In this work, we only use one ion and two motional mode. The other modes can be used for demonstration of three mode squeezed state. The internal electronic state of the ion is initialized to $|\!\!\downarrow\rangle$ by a combination of $\sigma^{+}$ polarized 397 laser and linearly polarized 866~nm, 854~nm lasers. We can discriminate internal state using fluorescence detection with the 397~nm cycling transition between $|S_{1/2}\rangle$ and $|P_{1/2}\rangle$ and an auxiliary 866~nm optial pumping transition\cite{detect}.

\subsection{Dissipation Process}
\begin{figure}
    \centering
    \includegraphics[width=\columnwidth]{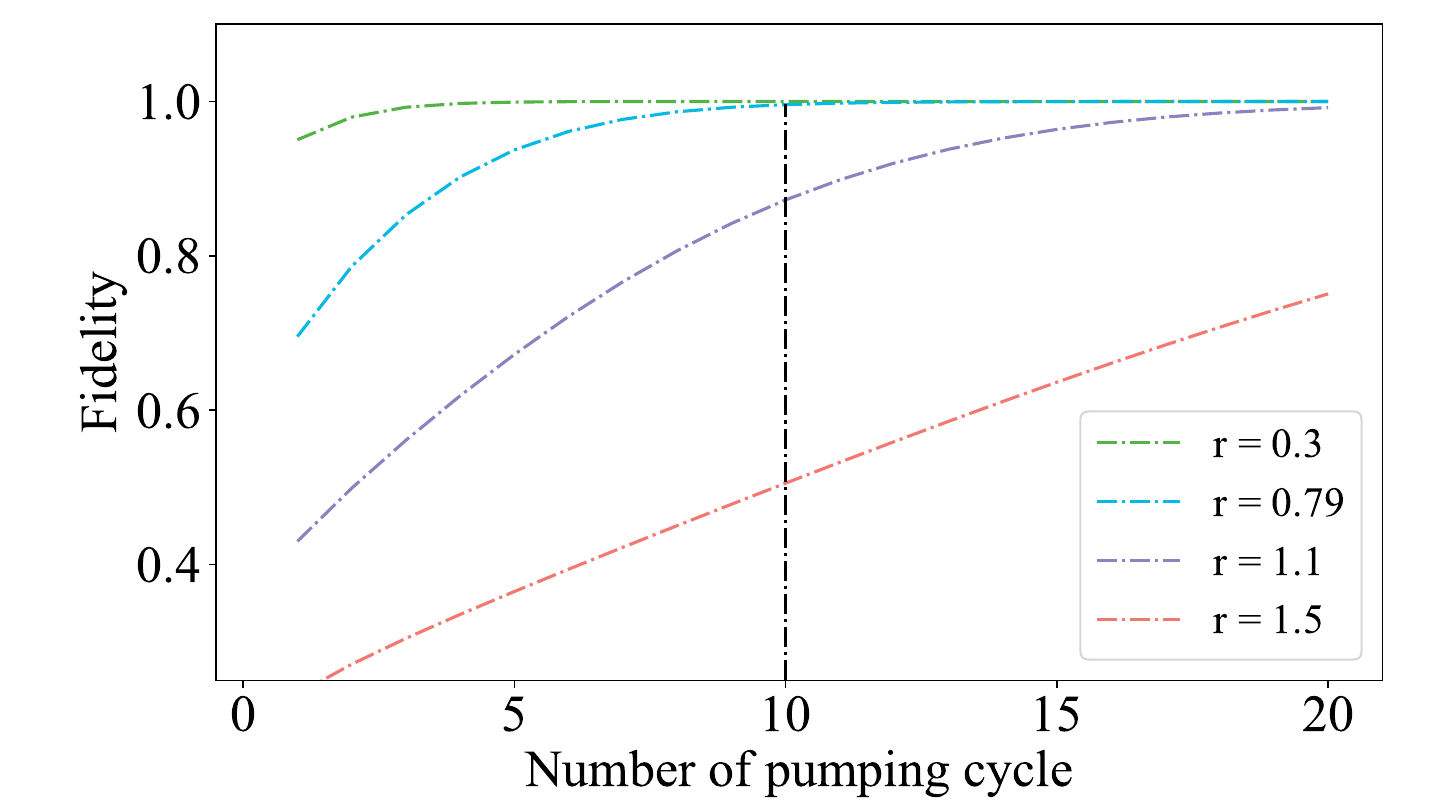}
    \caption{\textbf{The relationship between state fidelity and the number of dissipation cycle.} The simulated result of dissipative process with squeezed parameter $r=0.1,0.79,1.1$ and $1.5$. More pumping cycle is needed for larger squeezed parameter.}
    \label{fig:S3}
\end{figure}

In this work, we use a method based on reservoir engineering and demonstrate its effectiveness in generating TMSS between quantum harmonic oscillators. There are equally spaced energy eigenstates of the harmonic oscillator $|n\rangle$, which can be driven by creation and annihilation operators $a^{\dagger}$ and $a$. In trapped ions system, these operators can be used to cool the harmonic oscillators to ground states\cite{RevModPhys.75.281}. If we choose an engineered annihilation operator composed by a linear combination of creation and annihilation operators, which can be described by $\hat{K} = \hat{U} a \hat{U}^{\dagger}$, we always can find a set of energy eigenstates $\hat{U}|n\rangle$. As for multi-mode harmonic oscillators, we can choose a combination of multi-mode creation and annihilation operators described by $\hat{K_i} = \hat{U} a_i \hat{U}^{\dagger}$ corresponding to an energy eigenstates $\hat{U}|n_1,n_2,...n_i,...\rangle$. There exists a joint ground state $\hat{U}|0,0,...\rangle$ of these engineered creation and annihilation operators. In trapped ions experiment, we can cool in
this joint ground state by coupling the oscillators to an ancilla spin with Hamiltonian $\hat{H}_i^- = \hbar (\Omega_{i}\hat{K}_i\sigma^{+} +\Omega_{i}^{*} \hat{K}^{\dagger}_i \sigma^{-})$. When we add these Hamiltonian, the energy of the engineered states will decrease with spin flopping from $|\!\!\downarrow\rangle$ to $|\!\!\uparrow\rangle$, such as from $|\!\!\downarrow, n\rangle$ to $|\!\!\uparrow, n-1\rangle$.  Another spin optical pumping laser is introduced to generate dissipation, that pumps back the spin to $|\!\!\downarrow\rangle$. Thus, we just need to add these engineered spin-motion coupling and optical pumping by sequence, we can get the joint ground state finally. This process can start from a general initial state. For example, thermal states are easily obtained after Doppler cooling in trapped ions system, which can be used as the starting. We also do a numerical simulation of the dissipation process, we find that the larger squeezed parameter, the more dissipation cycle is needed, as shown in Figure \ref{fig:S3}.  In our simulation, we use $U_{1,2} = exp(-iH_{1,2}\times \frac{\pi}{2\Omega_{1,2}})$ to describe the coherent part. Under this coherent control, the output state is $\rho_{out} = U_{1,2}\rho_{in}U_{1,2}^\dagger$ .  After the coherent part, we calculate the spin partial trace of the output state $\rho_{motion} = Tr_{spin}(\rho_{out})$ to simulate the pumping process. Then the input state of the next cycle can be written as $\rho_{in} = |\!\!\downarrow \rangle\langle\downarrow\!\!|\bigotimes \rho_{motion} $. We can consider the drift of motional frequency during the coherent control by $U_{1,2} = exp(-i(\hbar\Delta_1a_1a_1^\dagger + \hbar\Delta_2a_2a_2^\dagger + H_{1,2})\times \frac{\pi}{2\Omega_{1,2}})$.  We simulate the effect of drift through random sampling of $\Delta_1$ and $\Delta_2$. In this work, 10 cycles for $r=0.79$ is enough to reach the steady state.  It's worth pointing out that the increased sensitivity to the stability of the driven field frequency due to the more dissipation cycles is a limitation for the larger value squeezing fidelity.

\subsection{Original Data}
Here we show the original fitting results of the TMSS population in the Fock basis and the engineered basis, which is shown in main text Figure 2.

\begin{figure}
    \centering
    \includegraphics[width=\columnwidth]{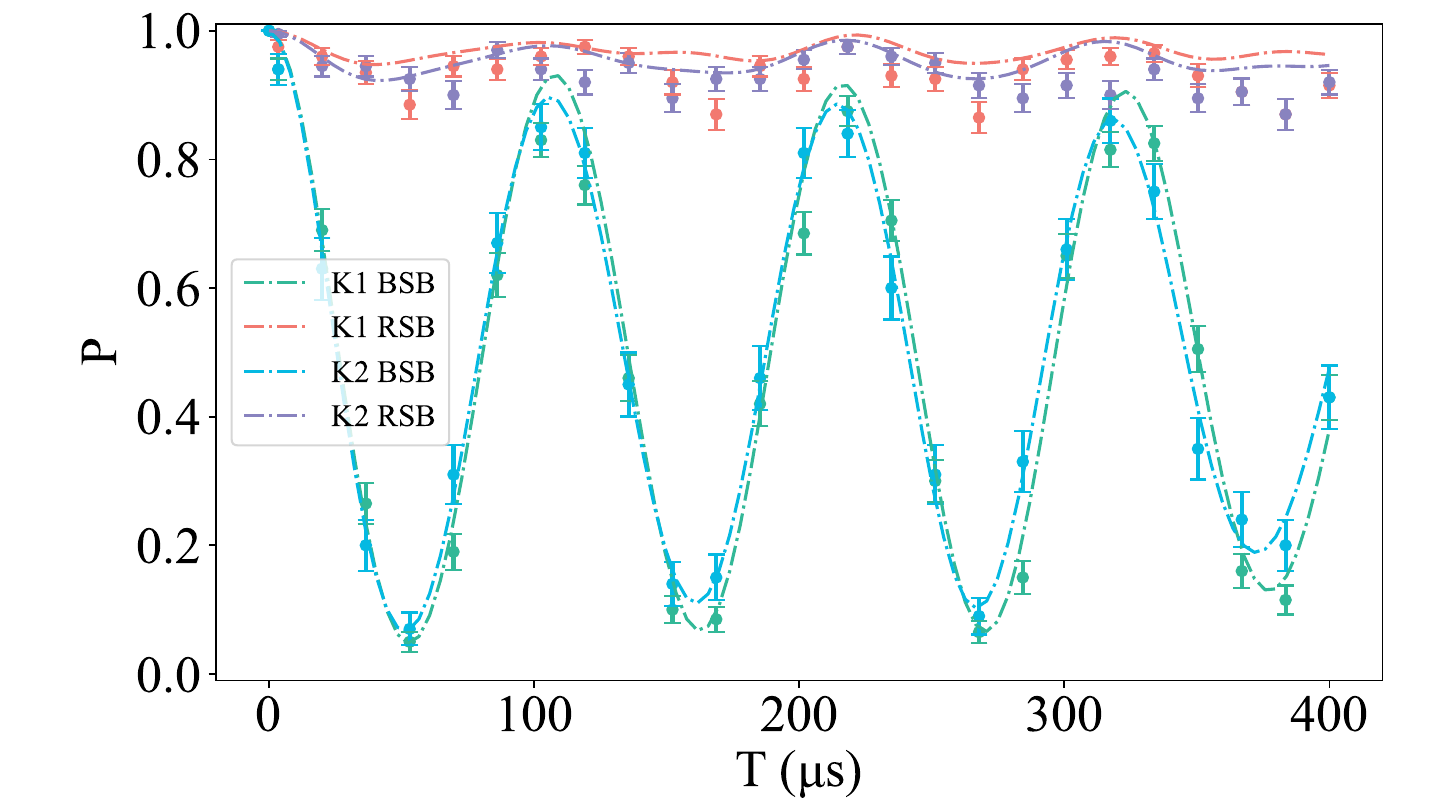}
    \caption{\textbf{State analysis in engineered basis.}The experimental data and fitting results using $\hat{H}^{+}_{1,2}$ as probe pulse, labeled with $\rm{K1,2~BSB}$. In contrast, applying $\hat{H}_{1,2}$ instead provide minimal change in the spin population, labeled with $\rm{K1,2~RSB}$, indicating the generation of a stabilized state. The parameters of fitting results are shown in Table S1. Each point is obtained from 200 repetitions. Error bars are the standard deviations of 200
repetitions.}
    \label{fig:S4}
\end{figure}

We characterize TMSS with two different methods. Firstly, we can obtain the fidelity after dissipative process by the fitting results of blue sideband Rabi flopping of Bogoliubov operators respectively with the Hamiltonian as: $\hat{H}^{+}_{1,2} = \hbar \Omega(\hat{K}_{1,2}\sigma^{-} + \hat{K}_{1,2}^{\dagger} \sigma^{+})$. The population of the state $|\downarrow\rangle$ can be written as a function of the blue sideband pulse duration t:
\begin{equation}
    P_{1,2}(|\downarrow\rangle) = \frac{1}{2} \sum_{n}p_{1,2}(n)(1+e^{-\gamma_nt}cos(\Omega_{n}t)),
    \label{BSB-vs-t}
\end{equation}
where $p_K(n)$ is the population of the nth energy eigenstate, $\Omega_{n}$ is the Rabi rate for the transtion between $|\downarrow\rangle, \hat{U}|n\rangle$ and $|\uparrow\rangle, \hat{U}|n+1\rangle$ and $\gamma_n$ describe the decoherence of ions and laser. We can obtain the population $p_{1,2}(n)$ by fitting the formula Eq.~\ref{BSB-vs-t}, as shown in Figure~S4. According to the fitting results, we can obtain the lower bound of state fidelity. $P_K(n,m)$ is used to describe the population of state $|n,m\rangle$.  We introduce $A=P_K(0,0)$, $B = \sum_{i\ge0} P_K(i,0)$, $C = \sum_{j\ge0} P_K(0,j)$, and $D = \sum_{i,j\ge1} P_K(i,j)$ to prove that $A\ge p_1(0)p_2(0)$, where $p_1(0) = A+B$, $p_2(0)=A+C$ and $A+B+C+D = 1$. Equivalently, it means $A-(A+B)(A+C)\ge0$. We can  simplify the equation to $AD-BC\ge 0 \leftrightarrow AD-(p_1(0)-A)(p_2(0)-A)\ge 0$, which means we need to analyse this inequation $-A^2 + (p_1(0)+p_2(0)+D)A-p_1(0)p_2(0)\ge0$. From our fitting results, $A+B = p_1(0)=0.91$, $A+C = p_2(0) = 0.95$, $C+D = 0.09$ and $B+D=0.05$ (from this we can obtain $A>0.86,B<0.05,C<0.09$ ). Under this constraint,  this inequation is ture. So we can use the $p_1(0)p_2(0)$ to describe the lower bound of fidelity.

\begin{figure}
    \centering
    \includegraphics[width=\columnwidth]{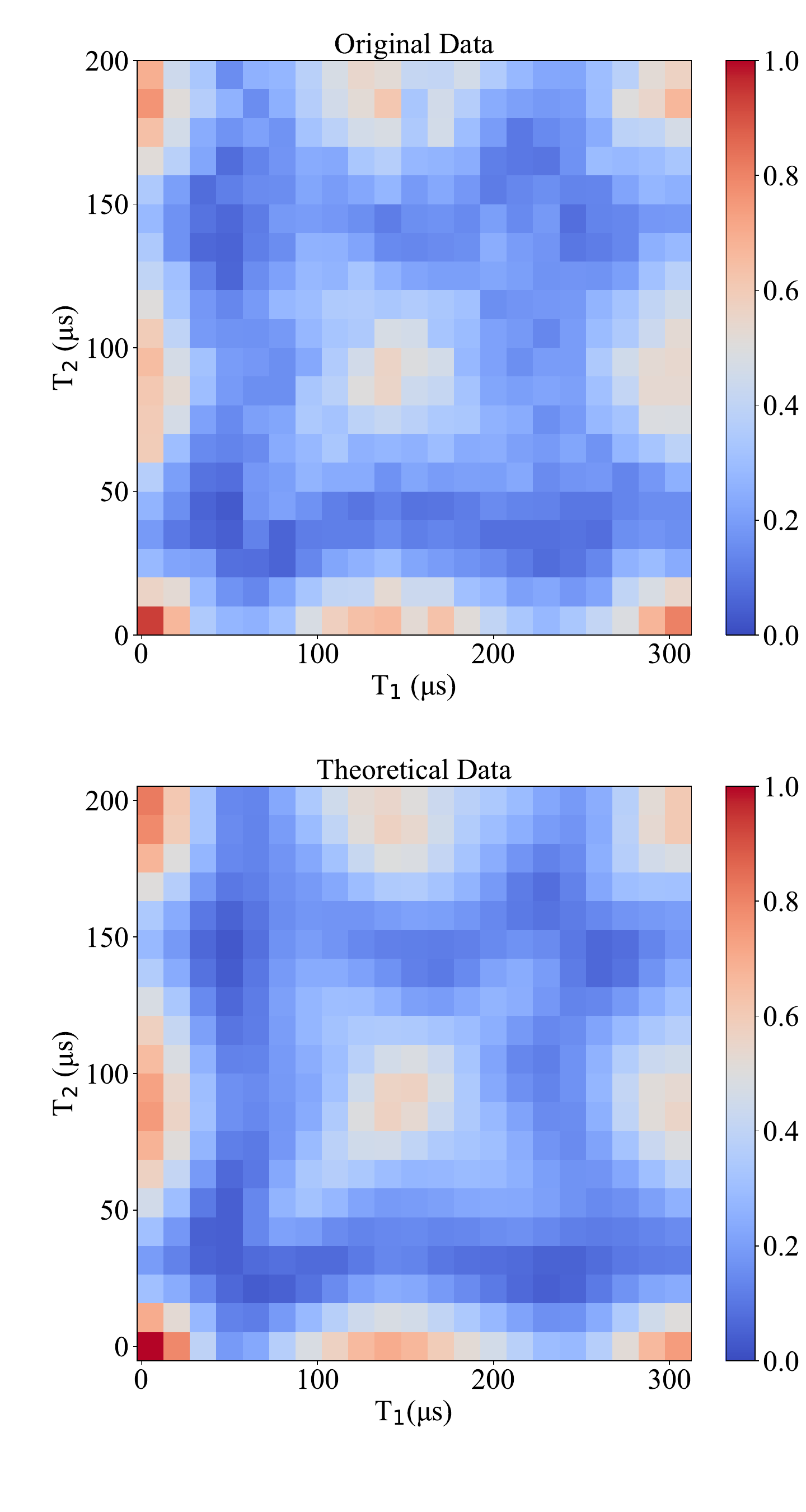}
    \caption{\textbf{State analysis in Fock basis.}The experimental data and fitting result using the blue sideband transition in Fock basis as probe pulse. Panel \textbf{a} is the experimental data and panel \textbf{b} is the theoretical result. The parameters of fitting results are shown in Table S1.}
    \label{fig:S5}
\end{figure}

We also use another method to verify these states independently. For TMSS, we use three internal state ($|\!\!\downarrow\rangle$, $|\!\!\uparrow\rangle$, and $|\rm{AUX}\rangle \equiv |L=2, J=5/2, M_J = + 5/2\rangle$) to measure the populations of the energy eigenstates. We perform blue sideband transtion $|\!\!\downarrow, n_1\rangle \leftrightarrow |\!\!\uparrow, n+1_1\rangle$ and $|\!\!\downarrow, n_2\rangle \leftrightarrow |\rm{AUX}, n+1_2\rangle$ in sequence with the Blue sideband Hamiltonian as:$\hat{H}_{1,2}^+ = \hbar \Omega(\hat{a}_{1,2}\sigma^{-} + \hat{a}^{\dagger}_{1,2} \sigma^{+})$. The population of the state $|\!\downarrow\rangle$ can be written as a function of duration $t_1$ and $t_2$ of the two blue sideband pulse respectively:
\begin{equation}
\begin{aligned}
     P(|\!\!\downarrow\rangle) = \frac{1}{2} \sum_{n,m}p(n,m)&(1+e^{-\gamma_nt_1}cos(\Omega_{n}t_1))\\
    &(1+e^{-\gamma_mt_2}cos(\Omega_{m}t_2))
\end{aligned}
    \label{BSB2}
\end{equation}
where $p(n,m)$ is the population of the $n$-th energy eigenstate of mode 1 and $m$-th energy eigenstate of mode 2. We can obtain the population $p(n,m)$ by fitting each set of data by the Eq.~\ref{BSB2}. The original data with squeezed parameter $r = 0.79$ are shown in Figure \ref{fig:S5}. The population fitting results are shown as Table S2.

\renewcommand\thetable{S\arabic{table}}
\begin{table}[h]
    \renewcommand{\arraystretch}{2}
    \centering
    \begin{tabular}{|c|c|c|c|c|c|}
        \hline
        n & 0 & 1 & 2 & 3 & 4 \\
        \hline
        population in $K_1$ basis & 0.91(3) & 0.03(2) & 0.03(2) & 0.01(2) & 0.02(2) \\
        \hline
        population in $K_2$ basis & 0.95(3) & 0.03(2) & 0.02(2) & 0.006(18) & 0.009(18) \\
        \hline
        
    \end{tabular}
    \caption{Fitting results for data using $\hat{H}^{+}_{1,2}$ as probe pulse, shown in Figure S4. Equation \ref{BSB-vs-t} is used for the fitting.}

    \label{tab:bsbel}
\end{table}

\begin{table}[h]
\renewcommand{\arraystretch}{2}
    \centering
    \begin{tabular}{|c|c|c|c|c|c|}
        \hline
        State & $|0,0\rangle$ & $|0,1\rangle$ & $|0,2\rangle$ & $|0,3\rangle$ & $|0,4\rangle$ \\
        \hline
        Population &\textbf{ 0.53(2)} & 0.02(1) & 5E-4(1E-2) & 1E-3(1E-2) & 0.02(1) \\
        \hline
         & $|1,0\rangle$ & $|1,1\rangle$ & $|1,2\rangle$ & $|1,3\rangle$ & $|1,4\rangle$ \\
        \hline
         & 0.03(1) &\textbf{ 0.23(1)} & 0.03(1) & 3E-4(1E-2) & 0.01(1) \\
        \hline
         & $|2,0\rangle$ & $|2,1\rangle$ & $|2,2\rangle$ & $|2,3\rangle$ & $|2,4\rangle$ \\
        \hline
         & 0.001(14) & 0.02(1) & \textbf{0.10(1) }& 6E-5(1E-2) & 1E-4(1E-2) \\
        \hline
         & $|3,0\rangle$ & $|3,1\rangle$ & $|3,2\rangle$ & $|3,3\rangle$ & $|3,4\rangle$ \\
        \hline
         & 0.01(1) & 1E-4(1E-2) & 0.01(1) & \textbf{0.06(1)} & 2E-4(1E-2) \\
        \hline
         & $|4,0\rangle$ & $|4,1\rangle$ & $|4,2\rangle$ & $|4,3\rangle$ & $|4,4\rangle$ \\
        \hline
         & 0.01(1) & 2E-4(1E-2) & 5E-5(1E-2) & 3E-4(1E-2) & \textbf{0.03(1)} \\
        \hline
        
    \end{tabular}
    \caption{ Fitting results for data as shown in Figure S5. Equation \ref{BSB2} is used for the fitting.}
    \label{tab:bsbfock}
\end{table}

\subsection{Three mode squeezed state}
For the three mode squeezed states case, we use three collective vibrational modes in radial directions of a two ions chain and the internal state of two trapped ions as the reservoir. The three mode squeezed states is defined as $\hat{S}(r)|0,0,0\rangle$, where
\begin{equation}
\begin{aligned}
    S(r) = exp[r (&a_1a_2 + a_2a_3 + a_3a_1 - \\
    &a_1^{\dagger}a_2^{\dagger}-a_2^{\dagger}a_3^{\dagger}-a_3^{\dagger}a_1^{\dagger})]\\
\end{aligned}
\end{equation}
is the two mode squeezing operator with squeezing parameter $r$. With three mode squeezing operator, we can obtain the three mode Bogoliubov operators:
\begin{equation}
\begin{aligned}
    \hat{K}_{j=1,2,3} =& \hat{S}(r)a_{j}\hat{S}(r)^{\dagger} 
    = \sum_{i=1}^{3} (f_i^j a_i + g_i^j a^{\dagger}_i)
\end{aligned}
\end{equation}
where  $f_i^j$ and $g_i^j$ are functions of the squeezed parameters $r$. The formulae of these functions are written as:

\begin{eqnarray}
    f_1^1 = (2cosh(r)+cosh(2r))/3,\nonumber\\
    g_1^1 = (2sinh(r)+sinh(2r))/3,\nonumber\\
    f_2^1 = (-cosh(r)+cosh(2r))/3,\nonumber\\
    g_2^1 = -(sinh(r)+sinh(2r))/3,\nonumber\\
    f_2^1 = f_3^1 = f_1^2 = f_3^2 = f_1^3 = f_2^3,\nonumber\\
    f_1^1 = f_1^2 = f_1^3 ,   g_1^1 = g_1^2 = g_1^3,\nonumber\\
    g_2^1 = g_3^1 = g_1^2 = g_3^2 = g_1^3 = g_2^3.
\end{eqnarray}
Same as before, the three mode squeezed state is the joint ground state of the three Bogoliubov
operators. We can prepare the three mode squeezed states by applying three mode dissipative process respectively with the engineered spin-motion coupling Hamiltonians as:
\begin{equation}
    \hat{H}_{1,2,3}^- = \hbar\Omega (\hat{K}_{1,2,3}\sigma^{+} + \hat{K}_{1,2,3}^{\dagger}\sigma^{-}),
\end{equation}

\begin{figure}[t]
    \centering
    \includegraphics[width=\columnwidth]{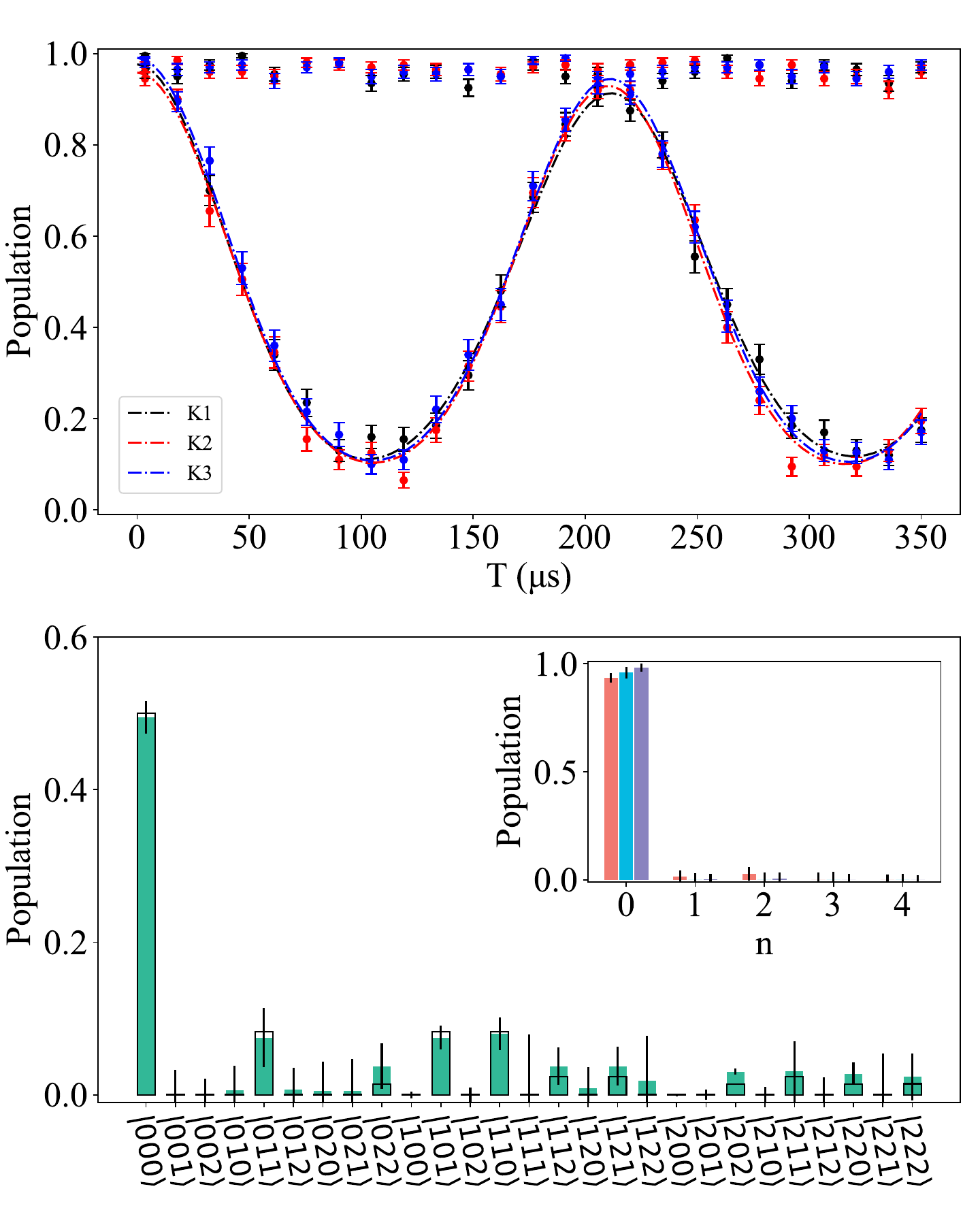}
    \caption{\textbf{Experimental results of three mode squeezed states.} \textbf{a,} Blue sideband transition in the engineered basis.   \textbf{b,} Population fitting results in Fock basis.The blue bars show the experimental results and the black frame shows the population of ideal three mode squeezed stats. The inset figure shows the population fitting results in engineered basis.}
    \label{fig:S6}
\end{figure}

We also use the blue sideband Rabi flopping of Bogoliubov operators to obtain the fidelity of the states. The rabi flopping with two ions and the fitting results are shown in Figure S6.
The population of the state $|\!\downarrow\downarrow\rangle$ can be written as a function of the blue sideband pulse duration t
\begin{equation}
    P_{1,2,3}(|\!\downarrow\downarrow\rangle) = \sum_{n}p_{1,2,3}(n) e^{-\gamma_nt}\frac{cos^2(\frac{1}{2}\sqrt{\Omega^2_{n}+\Omega^2_{n+1}}t)}{(\Omega^2_{n}+\Omega^2_{n+1})^2}.
    \label{BSB-two-vs-t}
\end{equation}
We can obtain the population $p_{1,2,3}(n)$ by fitting each set of data by the Eq.\ref{BSB-two-vs-t}. The lower bounds of fidelity of the state is $87(6)\%$.
We also verify the population in Fock states basis. For three mode squeezed states, we need three transitions to obtain the all population information. Specifically, 
$|\!\! \downarrow \rangle$, $|\!\! \uparrow \rangle$, $|\rm{AUX}_1\rangle$, and $|\rm{AUX}_2\rangle \equiv |L=2, J=5/2, M_J = + 3/2\rangle$ are employed. We perform three blue sideband transtions $|\!\!\downarrow, n\rangle \leftrightarrow |\!\!\uparrow, n+1\rangle$, $|\!\!\downarrow, m\rangle \leftrightarrow |\rm{AUX}_1, m+1\rangle$, and $|\!\!\downarrow, l\rangle \leftrightarrow |\rm{AUX}_2, l+1\rangle$  in sequence.
The population of the state $|\!\downarrow\downarrow\rangle$ can be written as a function of duration $t_1$ , $t_2$and $t_3$ of the three blue sideband pulse respectively:
\begin{eqnarray}
    P(|\!\downarrow\downarrow\rangle) &=& \sum_{n}p_{1,2,3}(n,m,l) e^{-(\gamma_nt_1+\gamma_mt_2+\gamma_lt_3)}\nonumber\\
    &&\frac{cos^2(\frac{1}{2}\sqrt{\Omega^2_{n}+\Omega^2_{n+1}}t_1)}{(\Omega^2_{n}+\Omega^2_{n+1})^2} \nonumber \\
    &&\frac{cos^2(\frac{1}{2}\sqrt{\Omega^2_{m}+\Omega^2_{m+1}}t_2)}{(\Omega^2_{m}+\Omega^2_{m+1})^2} \nonumber\\
    &&\frac{cos^2(\frac{1}{2}\sqrt{\Omega^2_{l}+\Omega^2_{l+1}}t_3)}{(\Omega^2_{l}+\Omega^2_{l+1})^2},
    \label{BSB-vs-t1t2t3}
\end{eqnarray}
where $p(n,m,l)$ is the population of the $n$-th energy eigenstate
of mode 1, $m$-th energy eigenstate of mode 2 and $l$-th energy eigenstate of mode 3. We can obtain the population p(n, m) by fitting each set of data by Eq.\ref{BSB-vs-t1t2t3}. The population fitted in Fock state basis and Bogoliubov basis are all shown in Figure S6. To verify the phase coherence, we check the inequality\cite{mCVCriterion} $\Delta_{\rm{EPR}} = \langle X_+^2\rangle + \langle P_-^2\rangle < 1/2$, where $X_{\pm} = X_{1} \pm \frac{1}{\sqrt{2}}(X_{2} + X_{3})$ and $P_{\pm} = P_{1} \pm \frac{1}{\sqrt{2}}(P_{2} + P_{3})$. In this work, we obtain $\Delta_{\rm{EPR}} = 0.22(2) < 0.5$ with squeezed parameter $r=0.5$ after 10 dissipation cycles, which is close to the ideal three mode squeezed state result$\Delta_{\rm{EPR}} = 0.208$. Three mode squeezed state could be utilized for the estimation of three parameters simultaneously. We can choose three reciprocal axes, such as $X_1+X_2$, $X_1+X_3$, and $P_1-P_2-P_3$. According to the numerical simulation, we could achieve 4.0 dB, 4.0 dB, and 7.7 dB improvement over the quantum limit along these three axes respectively.

\end{document}